\documentclass[conference]{IEEEtran}

\usepackage{todonotes}
\usepackage{graphicx}
\graphicspath{{../pdf/}{../jpeg/}}
\DeclareGraphicsExtensions{.pdf,.jpeg,.png}

\usepackage{breqn}

\usepackage[caption=false]{subfig}

\usepackage{textcomp}

\ifCLASSINFOpdf
\else
\fi
\usepackage{tikz}
\usepackage{lipsum}

\newcommand\copyrighttext{%
  \footnotesize \textcopyright 2019 IEEE.  This paper is published in the Proceedings of IEEE LCN 2019. Personal use of this material is permitted.  Permission from IEEE must be obtained for all other uses, in any current or future media, including reprinting/republishing this material for advertising or promotional purposes, creating new collective works, for resale or redistribution to servers or lists, or reuse of any copyrighted component of this work in other works.}
\newcommand\copyrightnotice{%
\begin{tikzpicture}[remember picture,overlay]
\node[anchor=south,yshift=10pt] at (current page.south) {\fbox{\parbox{\dimexpr\textwidth-\fboxsep-\fboxrule\relax}{\copyrighttext}}};
\end{tikzpicture}%
}

\begin{document}
%
\title{A Framework for Multiaccess Support for Unreliable Internet Traffic using Multipath DCCP}

\author{\IEEEauthorblockN{Markus Amend, Eckard Bogenfeld}
\IEEEauthorblockA{Deutsche Telekom\\
Darmstadt, Germany\\
markus.amend@telekom.de}
\and
\IEEEauthorblockN{Milan Cvjetkovic, Veselin Rakocevic}
\IEEEauthorblockA{City, University of London\\
United Kingdom\\
veselin.rakocevic.1@city.ac.uk}
\and
\IEEEauthorblockN{Marcus Pieska, Andreas Kassler\\ and Anna Brunstrom}
\IEEEauthorblockA{Karlstad University, Sweden\\
andreas.kassler@kau.se}}


%


\maketitle

\begin{abstract}
Mobile nodes are typically equipped with multiple radios and can connect to multiple radio access networks (e.g. WiFi, LTE and 5G). Consequently, it is important to design mechanisms that efficiently manage multiple network interfaces for aggregating the capacity, steering of traffic flows or switching flows among multiple interfaces. While such multi-access solutions have the potential to increase the overall traffic throughput and communication reliability, the variable latencies on different access links introduce packet delay variation which has negative effect on the application quality of service and user quality of experience. In this paper, we present a new IP-compatible multipath framework for heterogeneous access networks. The framework uses Multipath Datagram Congestion Control Protocol (MP-DCCP) - a set of extensions to regular DCCP - to enable a transport connection to operate across multiple access networks, simultaneously. We present the design of the new protocol framework and show simulation and experimental testbed results that (1) demonstrate the operation of the new framework, and (2) demonstrate the ability of our solution to manage significant packet delay variation caused by the asymmetry of network paths, by applying pluggable packet scheduling or reordering algorithms.\end{abstract}


\renewcommand\IEEEkeywordsname{Keywords}


%
\IEEEpeerreviewmaketitle

\copyrightnotice

\section{Introduction}

The potential of multi-connectivity in network access solutions continues to gather interest with operators. The increased capacity provided by simultaneous deployment of different radio and fixed access networks (WiFi, LTE, 5G, DSL) offers the potential for improved service offering to home and business users, particularly for bandwidth-hungry live and streaming multimedia applications.

At the same time, to fully utilise the potential of multipath access, it is necessary to solve several important technical challenges, with packet delay variation being one of the most important. Reliable data delivery requires synchronisation of congestion and flow control mechanisms across different paths, and the asymmetry of path latencies presents a particular challenge for the real-time multimedia applications and for any latency-sensitive transport protocol or application. Currently, multimedia applications use a range of solutions to deal with the inherent bandwidth and delay variation in the network \cite{c1}, including layered video coding and numerous adaptation mechanisms, but a limitation in terms of how much variation the applications can efficiently handle has been well documented \cite{c2}. For multipath access to be used for effective traffic distribution, a solution must be found to manage the delay variation challenge, while at the same time making sure that this does not reduce the capacity benefit provided by multiple paths. 

There are several standardisation activities  for multi-connectivity solutions. The Broadband Forum recently developed a standard for Hybrid Access \cite{c3}, which defines a framework that enables converged network operators to offer their fixed access subscribers coordinated and simultaneous use of fixed broadband and 3GPP access networks. In 3GPP, the ATSSS study group (Access Traffic Steering Switching Splitting \cite{c4}) is specifying a flexible framework for the selection of access technologies, overcoming the single-access limitations of today’s devices and services. Steering (load balancing), switching (seamless handover), and splitting (capacity aggregation) are all being considered as critical in delivering the full benefits of multi-connectivity. However, access and path measurements are required for efficient traffic steering decisions, and this needs to be investigated further.

Over the last decade, IETF standardised the Multipath TCP protocol (MP-TCP \cite{c5}), which added path aggregation functionality to the classic TCP while inheriting many desirable properties, including the congestion control mechanism.  MP-TCP has been investigated in a number of research works (\cite{c6}\cite{c7}\cite{c8}\cite{c9} and many others), and has been demonstrated to deliver increased network capacity to applications with minimal impact on the data flow dynamics. MP-TCP is also considered by the 3GPP ATSSS study group as the solution for traffic splitting for applications requiring reliable TCP-based data delivery. Still, MP-TCP is limited to TCP services and excludes any other network protocol, in particular UDP. MP-TCP is based on a reliable in-order protocol, and thus it imposes these properties on any traffic that is transported via the solution. This is a problem for unreliable user traffic, causing unnecessary retransmissions and head-of-line blocking, leading to increased latency and delivery interruption. Due to the increasing share of UDP traffic, mainly caused  by the introduction of the QUIC protocol \cite{c10}, multi-access solutions must be able to efficiently cope with unreliable traffic such as  UDP. 

\begin{figure}[ht!]
\includegraphics[width=3.4in]{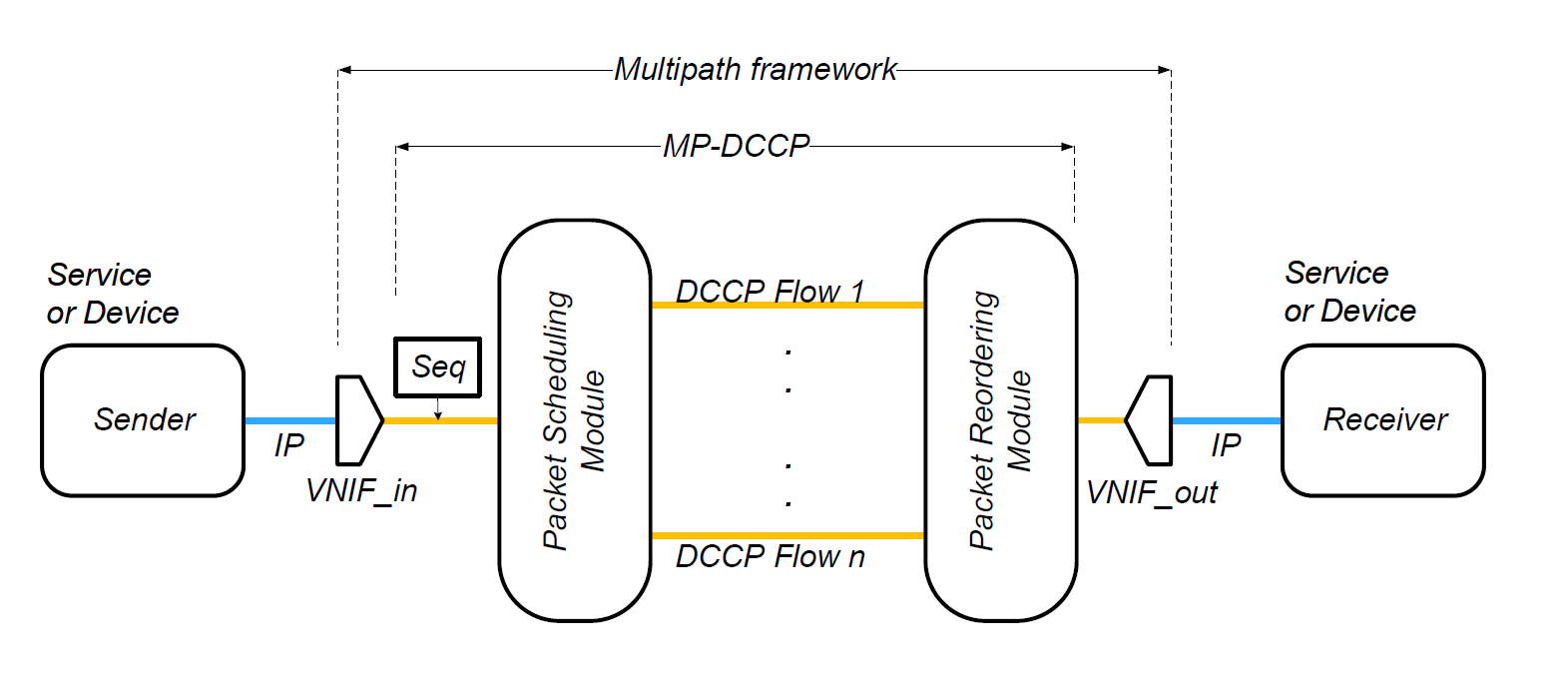}
\caption{Multipath Framework Architecture}
\label{fig:framework-architecture}
\end{figure}

The main contribution of this paper is the design and implementation of a new framework for multipath support for IP traffic. The new framework provides IP compatibility by integrating virtual network interfaces (VNIF) with a new protocol - Multipath Datagram Congestion Control Protocol (MP-DCCP), presented also in our recent IETF Internet Draft \cite{c10}. In the new framework, as presented in Fig.\ref{fig:framework-architecture}, the new MP-DCCP protocol is responsible for scheduling the traffic entering the VNIF\textsubscript{in} for transport over different radio or fixed access network paths. For this purpose, our framework includes a pluggable packet scheduling module which decides which IP packets to allocate to which DCCP flow. Optionally, a packet reordering module can be used to reorder packets at the exit point to remove latency variation introduced by heterogeneous paths. MP-DCCP  provides the network with path measurement capability, identified as the necessary ingredient for any effective traffic steering / switching / splitting solution. On each path, the traffic is encapsulated into separate DCCP flows. In general, MP-DCCP can be used either for all Internet traffic (TCP and non-TCP), or in parallel with MP-TCP, for the transport of non-TCP traffic. 

In this paper we present simulation and experimental results to both motivate and justify the presented framework. We use simulation to show that the throughput of RTP media streams on multiple paths with asymmetric latencies is significantly improved using our framework with the packet reordering module. We then present experimental results on a network testbed to demonstrate the positive impact of packet scheduling and packet reordering on the performance of the framework, and to demonstrate the ability of the framework to (a) remove packet delay variation from UDP streams, and (b) deal efficiently with sudden changes in path latencies.  

The new framework, the design of MP-DCCP, and principle design of pluggable scheduling and reordering algorithms are presented in more detail in section II. Section III presents the simulation and experimental setup and presents the performance evaluation results, demonstrating the benefits of the new framework. Finally, the paper is concluded in section IV.

\section{The New Multipath Framework}

The new IP compatible multipath framework is designed to operate in home access network scenarios where both fixed access network and a wireless access connection are available, or in a mobile scenario where multiple network interfaces are available (WiFi and LTE interface in most cases).

The development of the new multipath framework builds on a number of technical proposals for multi-connectivity support. One of the earliest concepts for aggregating heterogeneous access networks was \textit{IP bundling}, which is a Layer 3 approach first demonstrated by Deutsche Telekom in 2013 and currently sold as the \textit{Hybrid} product. The solution is capable of bundling fixed access lines (e.g. DSL) and mobile networks (e.g. LTE), providing customers with higher bandwidth and enhanced reliability. The DSL path is preferred whenever possible; all traffic that exceeds the DSL capacity is sent via the mobile network, which is referred to as a \textit{Cheapest Pipe First} approach. The IP bundling architecture assumes the DSL path to be static in bandwidth; this concept allows to aggregate network paths without performing path estimation. In the case of bottlenecks outside of the DSL link, this technique may be problematic: as there is no way of detecting the bottleneck, the path might be overloaded, causing packet loss. To address this shortcoming, our new framework provides path-based measurements.
MP-DCCP supports the new multipath framework by providing multiple DCCP tunnels between communicating points. The DCCP tunnels carry DCCP packet flows, ensuring traffic transport between the endpoints and allowing the framework to use DCCP functionality such as congestion control, RTT estimation and explicit connection establishment, providing this essential path-based information to handle traffic efficiently. At the same time, the use of MP-DCCP instead of a TCP-based solution provides support to unreliable, UDP-based traffic. Consequently, we identify two modes of operation for the framework: (1) it can be used as a stand-alone traffic bundling solution; (2) it can be configured to handle only UDP traffic and work in conjunction with a MP-TCP based solution which would handle TCP traffic.   

MP-DCCP is based on DCCP \cite{c12}, which was designed to provide different congestion control support for application traffic, together with unreliable, UDP-like support when application traffic does not require reliable delivery. DCCP is designed to provide full-duplex, connection-oriented data transfer, without provisioning for reliable in-order packet delivery or flow control. It provides congestion information for the end points, in a way similar to TCP. DCCP provides general support for various congestion control algorithms, in a modular fashion, enabling link quality estimation. The DCCP standard does not specify one single congestion control algorithm, rather several congestion control profiles have been defined for DCCP\cite{c13}\cite{c14}. There is a relatively limited body of academic research on the DCCP protocol, mostly investigating the differences between DCCP and TCP  \cite{c15} \cite{c16}, or performance of streaming video applications over DCPP \cite{c17} \cite{c18}. 

Within the new multipath framework, the DCCP tunnels serve two main purposes: (1) the integrated DCCP congestion control provides channel estimation in terms of path loss rate, bandwidth and round trip time; (2) the tunnels ensure redirection of traffic to the aggregation termination points. Layer 3 compatibility is achieved by employing virtual network interfaces to easily integrate in existing architectures such as home gateway. Virtual network interfaces are similar to their physical counterparts - the operator can configure settings such as the IP address or the netmask. The outgoing traffic is distributed to the DCCP tunnels by a scheduling algorithm, which chooses one of the DCCP sockets for each packet. At the receiving end, the solution implements modular packet reordering. A more detailed description of the packet flow can be found in our recent IETF Internet Draft \cite{c19}.


\subsubsection{Packet Scheduling}

The packet scheduling module distributes the data over the available DCCP flows. It can use the information from the DCCP flows, including the current estimate of the DCCP channel state, to make a scheduling decision. After the selection of a DCCP flow, the current packet to be scheduled for transmission is encapsulated into the flow.  

The scheduler itself can be designed to fulfill different objectives. Simple variants of the scheduling algorithm design may include the \textit{round robin} scheduler, where packets will be equally shared between the available paths; the \textit{fixed ratio} scheduler, where fixed weights are used to specify the ratio of packets scheduled on certain paths; or the \textit{cheapest pipe first} packet scheduler, where the operator can assign a cost value to each network path. The cheapest pipe first packet scheduler tries to minimize the overall cost by sending on cheaper paths whenever possible (i.e. as long as there is room in the congestion window). This scheduler is a simple and useful concept in the cases where the transmission on a line causes actual monetary cost, and as such is a commonly preferred concept for the operators.

In this work we analyse two well-known scheduling concepts: the \textit{smoothed RTT (SRTT)} scheduler, and the \textit{Out-of-order transmission for In-order arrival (OTIAS) }scheduler. The SRTT scheduler uses the RTT information provided by the DCCP congestion control to choose the currently available path that provides the lowest RTT. This simple concept is deployed in the MP-TCP reference implementation in the Linux Kernel. While it is known that this idea may lead to under-utilisation of the faster path \cite{c20}, we use it here as a simple example of a path-aware packet scheduling principle. 

The OTIAS scheduler \cite{c21} intentionally scrambles packets to compensate for asymmetric path properties.  Similar to SRTT, the OTIAS scheduler uses the estimated RTT values to choose the path that provides the shortest transmission time. In addition, the OTIAS scheduler estimates the time a packet would have to wait in the send queue of a socket and uses this time to deliberately overload flows with lower RTTs (i.e. schedule more packets than what a flow can currently send according to the congestion window). As the send queue builds up on the 'faster' path, the latency increases. Once the latency increases beyond the SRTT on the other path, OTIAS switches to this path. In the meantime, the queue at the first flow drains and the cycle repeats. The scheduler aims to ensure that  packets will arrive in the correct order at the receiver, thus reducing the need for a reordering mechanism.

\subsubsection{Packet Reordering}

The packet reordering module has an advantage over the packet scheduling module as it does not rely on average channel feedback to make scheduling decisions to prevent out-of-order packet delivery. 

Packet reordering, instead, is typically based on monitoring the received packets, buffering out-of-order packets and delaying their delivery to the application for a pre-defined timing threshold, expecting that during this time packets will arrive to fill the gaps. If the out-of-order packets do arrive within the timing threshold, they will be placed in their positions and ordered packet stream will be delivered to the application. If the out-of-order packets do not arrive within the timing threshold, the buffered packets will be forwarded to the application, which will have to deal with the consequence of any missing packets. This concept limits the buffering delay and presents an advantage in comparison to the MP-TCP-based framework, where for large latency variation, retransmissions may be requested for the delayed packets, resulting in increased overall delays for unreliable traffic.

Determination of the timing threshold requires two new protocol functionalities, which are introduced in MP-DCCP as two header options. The first is used to convey the RTT information to the receiving end of the framework. The second option is used to carry MP-DCCP packet sequencing information. To facilitate reordering, MP-DCCP deploys overall packet sequencing, which is overlayed on top of the standard sequencing used in DCCP.

Based on the use of the timing threshold, we identify three different reordering algorithms. In the \textit{static timing threshold} reordering, the timing threshold does not change, and is based on the difference in the RTTs for the two paths. If we label the path with larger latency 'slower' path, and the path with smaller latency 'faster' path, the static threshold can be calculated as \(\Delta RTT_{static} = RTT_{slower} - RTT_{faster}\). In the \textit{adaptive timing threshold} reordering, the value of the threshold is constantly recalculated based on the measured RTTs and RTT variations. For these two algorithms, the reordering module used the threshold to wait for out-of-order packets, and puts all packets received within the threshold into the correct order, based on the MP-DCCP packet sequencing. 

Finally, we use a third concept, \textit{delay equalisation}. In this algorithm, the measured timing threshold is also used, this time to delay the packets on the faster path in attempt to remove packet delay variation. This is done by introducing a buffer for each individual flow which delays traffic from the corresponding flow, ensuring that packets are delivered to the application in-order, regardless of when they arrive at the receiving end. In this case, packet sequencing is not used and all packets that arrive after the threshold are discarded.

\section{Performance Evaluation}
We use both simulation and experiments on a network testbed to evaluate the performance of the new framework. Firstly, we use simulation to address the following question: \textit{what is the nature of the interaction of the framework with congestion control of real-time UDP-based media applications}? After that, we use experiments on a network testbed to answer the following: \textit{what performance impact do the scheduling and reordering algorithms presented in section II have on the end-to-end performance?}  

\subsection{Simulation results}
In order to answer the first question, we evaluate the impact of our framework on UDP based real-time media applications that implement the NADA \cite{c26-nada} congestion control for RTP media streams \cite{c27-rmcat}. NADA is a congestion control that is currently being standardized to handle real-time UDP traffic. We use the network simulator ns3 \cite{c22-ns3} v3.29 with Direct Code Execution (DCE) \cite{c23-dce}, which uses the real Linux network stack of DCCP together with ns3. We use Linux Kernel version 4.1.0 from net-next-sim \cite{c24-net-next-sim} and the ns3 RMCAT  module \cite{c25-ns3-rmcat} that implements the NADA congestion control. 
In our simulation setup, we use a single video flow  being downloaded over two paths for 60 sec. The topology is similar to Fig. \ref{fig:framework-architecture}, except the receiver is a mobile node with two interfaces and the DCCP tunnel is terminated at the UE. The capacity of each link is not the bottleneck and the sending rate is limited at 1.5 Mbps given by the server video quality.

\begin{figure}[htp]

\centering
\subfloat[Adaptive reordering; DCCP-tunnels.]{%
\includegraphics[clip,width=0.98\columnwidth]{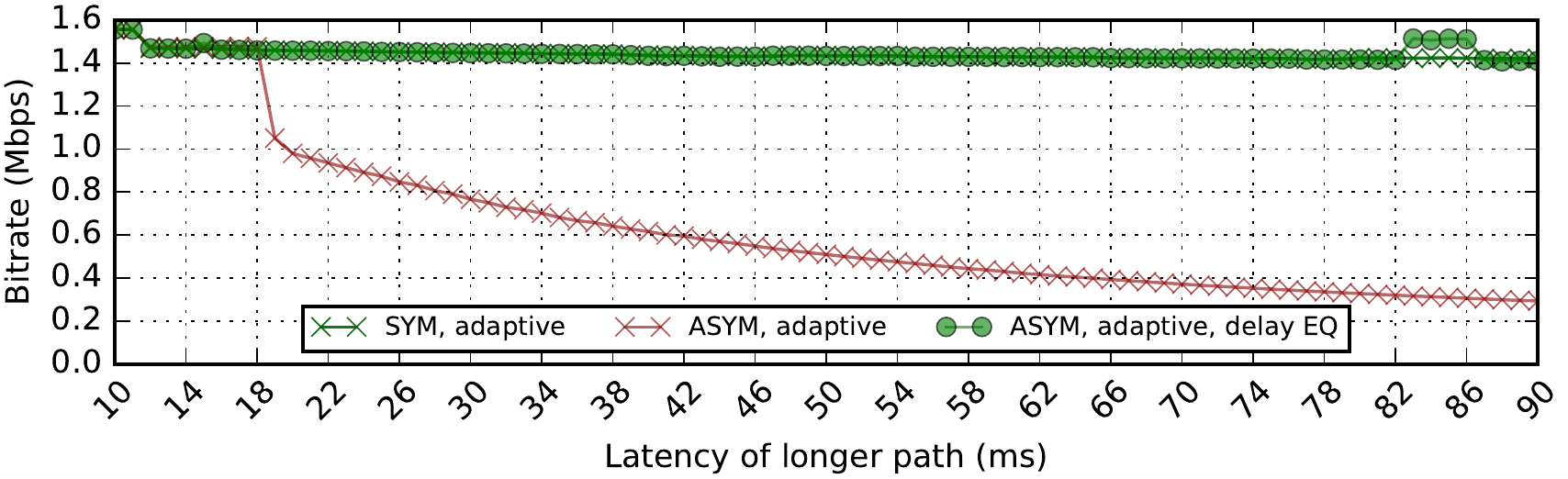}%
\label{fig:nada-dccp}
}

\centering
\subfloat[No reordering (with comparison); DCCP-tunnels.]{%
\includegraphics[clip,width=0.98\columnwidth]{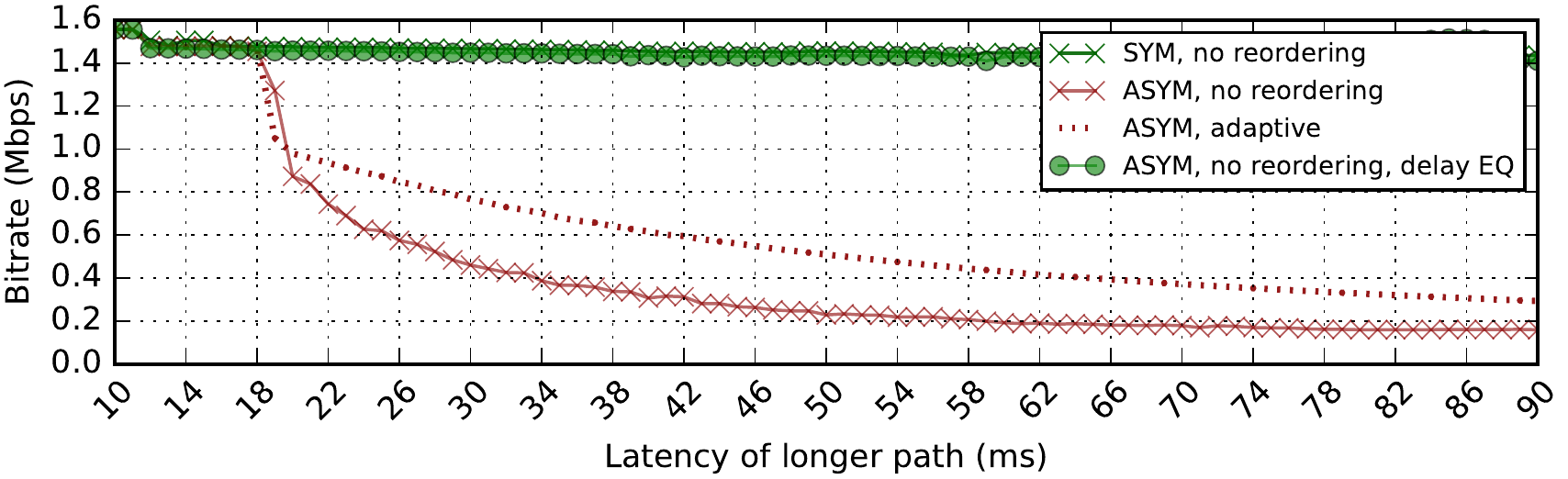}%
\label{fig:nada-dccp-pas}
}

\centering
\subfloat[Adaptive reordering; TCP-tunnels.]{%
\includegraphics[clip,width=0.98\columnwidth]{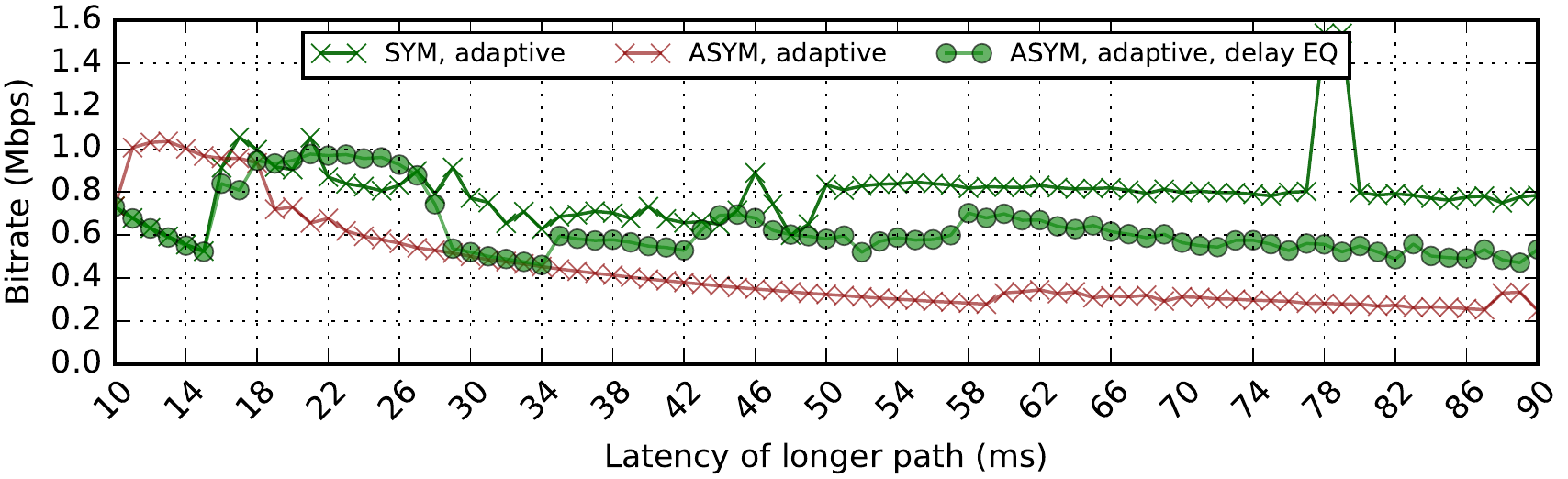}%
\label{fig:nada-tcp}
}

\caption{NADA/UDP throughput using a round-robin scheduler. In the symmetrical case (SYM), the latency applies to both paths; in the asymmetrical cases (ASYM), the latency applies to the 'slower' path, with the 'faster' path having a latency of 10 ms. \textit{Delay EQ} is short for delay equalisation; \textit{adaptive} denotes that adaptive reordering is used at the receiver. }
\label{fig:nada-multi}
\end{figure}

\begin{figure}[htp]
\centering
\includegraphics[clip,width=0.96\columnwidth]{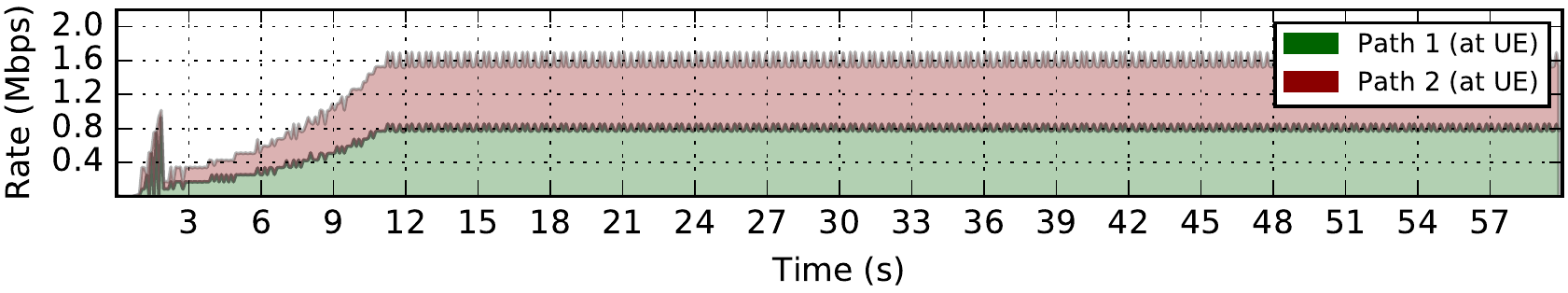}
\includegraphics[clip,width=0.96\columnwidth]{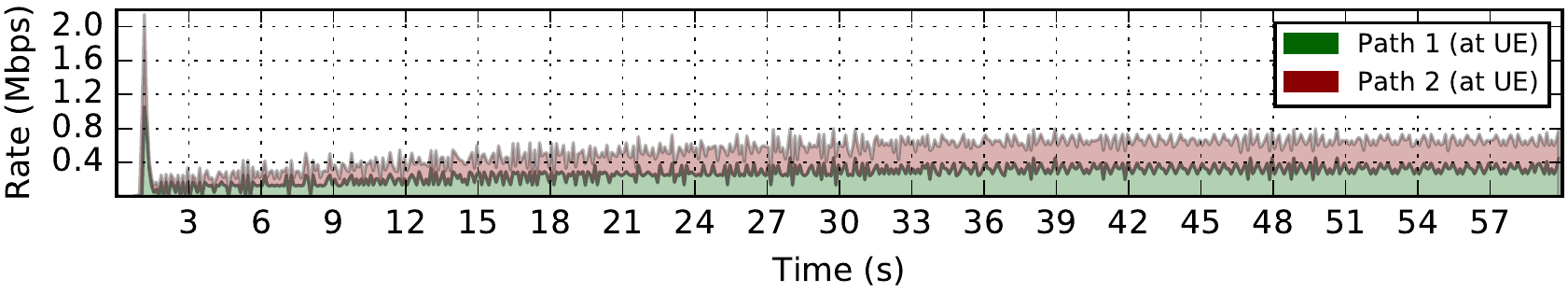}
\caption{Two examples pulled from Fig. \ref{fig:nada-multi}. In each case adaptive reordering and delay equalisation is used, with the latency being 10 and 90 ms. DCCP tunnels are used in the top figure; TCP tunnels are used in the bottom. }
\label{fig:nada-multi-ex}
\end{figure}


Fig. \ref{fig:nada-multi} shows the achievable rate of the video flow which is split over two paths using a round-robin scheduler and with either \textit{adaptive reordering} or \textit{no reordering} at the receiving side. We vary the path latency and compare three scenarios: symmetrical links, where the latency varies from 10 to 90 ms;  asymmetrical links, where the shorter path has a constant latency of 10 ms while the latency of the longer path varies between 10 and 90 ms; asymmetrical links but with activation of the delay equalization module at the receiving side (to expose a constant latency to the NADA congestion controller in the server). In Fig. \ref{fig:nada-dccp}, we are using adaptive reordering along with DCCP tunnels over the two paths, in Fig. \ref{fig:nada-dccp-pas}, we are using no reordering with DCCP tunnels, while in Fig. \ref{fig:nada-tcp} we use TCP tunnels with adaptive reordering to motivate the benefit of our approach.

As can be seen from Fig. \ref{fig:nada-dccp}, NADA has an adverse reaction to delay asymmetry. In our results, any asymmetry larger than 10 ms causes the NADA congestion control to reduce its sending rate. 
As one might expect, the introduction of delay equalization at the receiving side of the tunnel eliminates the adverse behavior. 

In Fig. \ref{fig:nada-dccp-pas}, we observe that  
the packet scrambling caused by no reordering seems to cause no problem as long as the links are symmetrical. When the links are asymmetric, delay equalization appears to solve the issue, with no need for further handling of any packet scrambling. The two red plots depict the outcome when there is asymmetry --- with the \textit{adaptive} case being borrowed from Fig. \ref{fig:nada-dccp} --- and illustrate that adaptive reordering helps marginally, but that it is not sufficient to achieve the application limited rate of 1.5 Mbps. Lastly, Fig. \ref{fig:nada-tcp} illustrates 
a significant performance degradation when using TCP-tunnels instead, showing the benefit of our approach. Fig. \ref{fig:nada-multi-ex} exemplifies how the maximum rate is achieved quickly when using DCCP-tunnels, whereas when we use TCP-tunnels the rate appears to stagnate at slightly more than a third of the maximal throughput.  


From these results we may conclude the following regarding how to handle NADA traffic: One needs not be unduly concerned with packet reordering, as long as the link asymmetry is rather small. On the other hand, providing a consistent RTT for the sending application appears to be more important. For that reason, TCP may be unsuitable as a tunneling solution for NADA traffic. While NADA is only a single congestion control solution among many others that UDP based applications may leverage, other delay-based congestion controls may experience similar problems. Furthermore, having a modular tunneling solution also allows for flexible deployment of schedulers and reordering mechanisms. Therefore, our framework can easily embrace future more advanced UDP congestion control algorithms.

\subsection{Experimental Testbed evaluation}

\begin{figure}[ht!]
\includegraphics[width=3.4in]{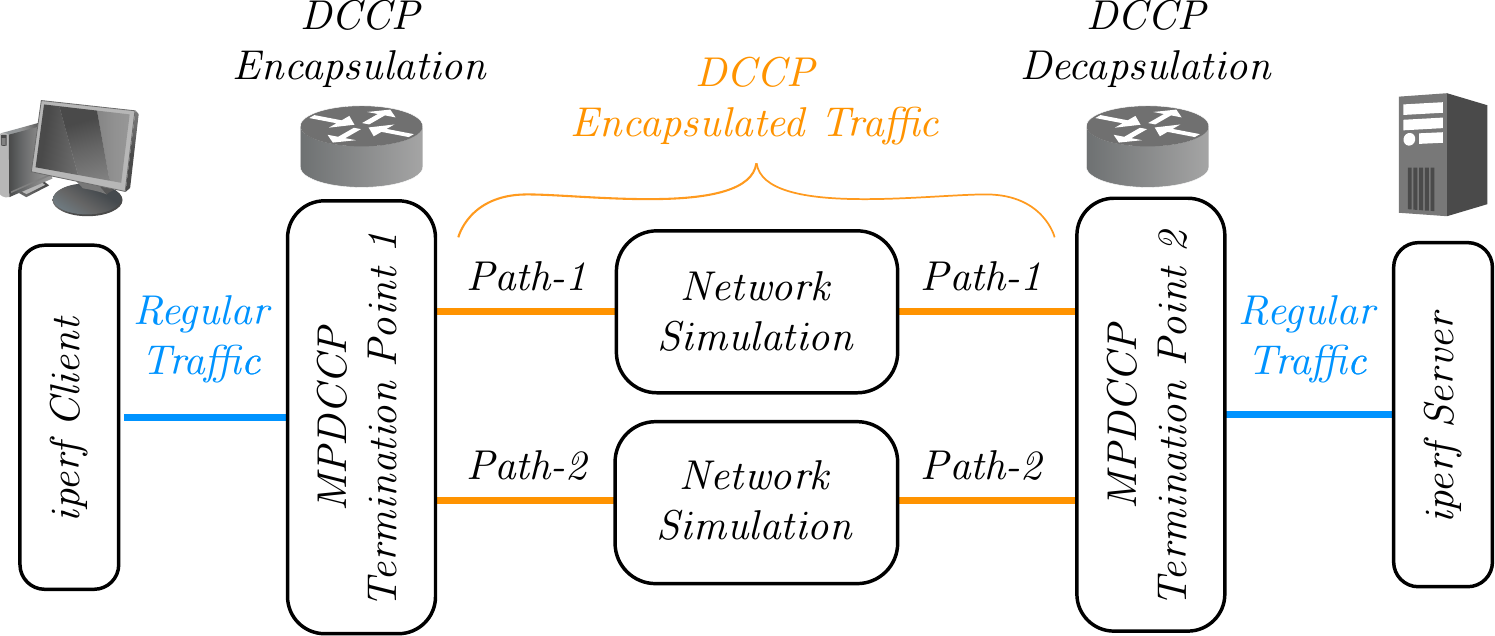}
\caption{Network Testbed Architecture}
\label{fig:testbed-architecture}
\end{figure}

To evaluate the performance of the new framework further, a prototype and network testbed have been developed. Fig \ref{fig:testbed-architecture} shows the architecture of the network testbed, which was developed on the basis of the framework architecture given in Fig. \ref{fig:framework-architecture}. In the testbed, the MP-DCCP \textit{bundling} solution is deployed between two MP-DCCP Terminating points (Fig. \ref{fig:testbed-architecture}), aggregating two network paths: \textit{Path-1 }and \textit{Path-2}. The endpoints iperf Client and iperf Server, which generate test traffic using the iperf network test tool, are connected to the third interface of each bundling endpoint. Additionally, there is a dedicated host on each path for the simulation of the access links. 

The  architecture and the access link simulation hosts are based on the PCengines APU board 1D4 embedded platform, equipped with a 1GHz AMD Bobcat dual-core processor1, 4GB of RAM and three Gigabit network interfaces. 

The MP-DCCP solution has been implemented as a kernel module in the Linux kernel 4.4, (supporting up to 4.19) as a proof-of-concept based on the current DCCP implementation in the Linux kernel \cite{c28} and the CCID2 congestion control algorithm \cite{c13}. The MP-DCCP source code consists of mandatory source files that are required for basic MP-DCCP operation, and optional loadable modules which can be loaded or unloaded during a MP-DCCP session. Many of the concepts used in MP-DCCP originate from the MP-TCP project and have been verified extensively in terms of stability, performance and usability.

\subsubsection {Performance Evaluation - Scheduling}

To evaluate the performance of the scheduling algorithms, experiments with a fixed-rate data stream of 1Mbps have been performed. 

The operation of the SRTT scheduler can be illustrated by increasing the latency of the preferred path during the transmission. As soon as the new latency value exceeds that of another network path, this network path will become the new preferred path. Two network paths have been set up to provide symmetric bandwidths of 10 Mbps. Additionally, the first path has been configured to 10 ms, and the second path to 20 ms of latency. In this scenario, the SRTT scheduler selects the first path. After 15 seconds, the latency on the first path has been increased from 10ms to 100 ms, forcing the scheduler to switch to the second path.

\begin{figure}[ht!] 
\centering
\includegraphics[width=3.3in]{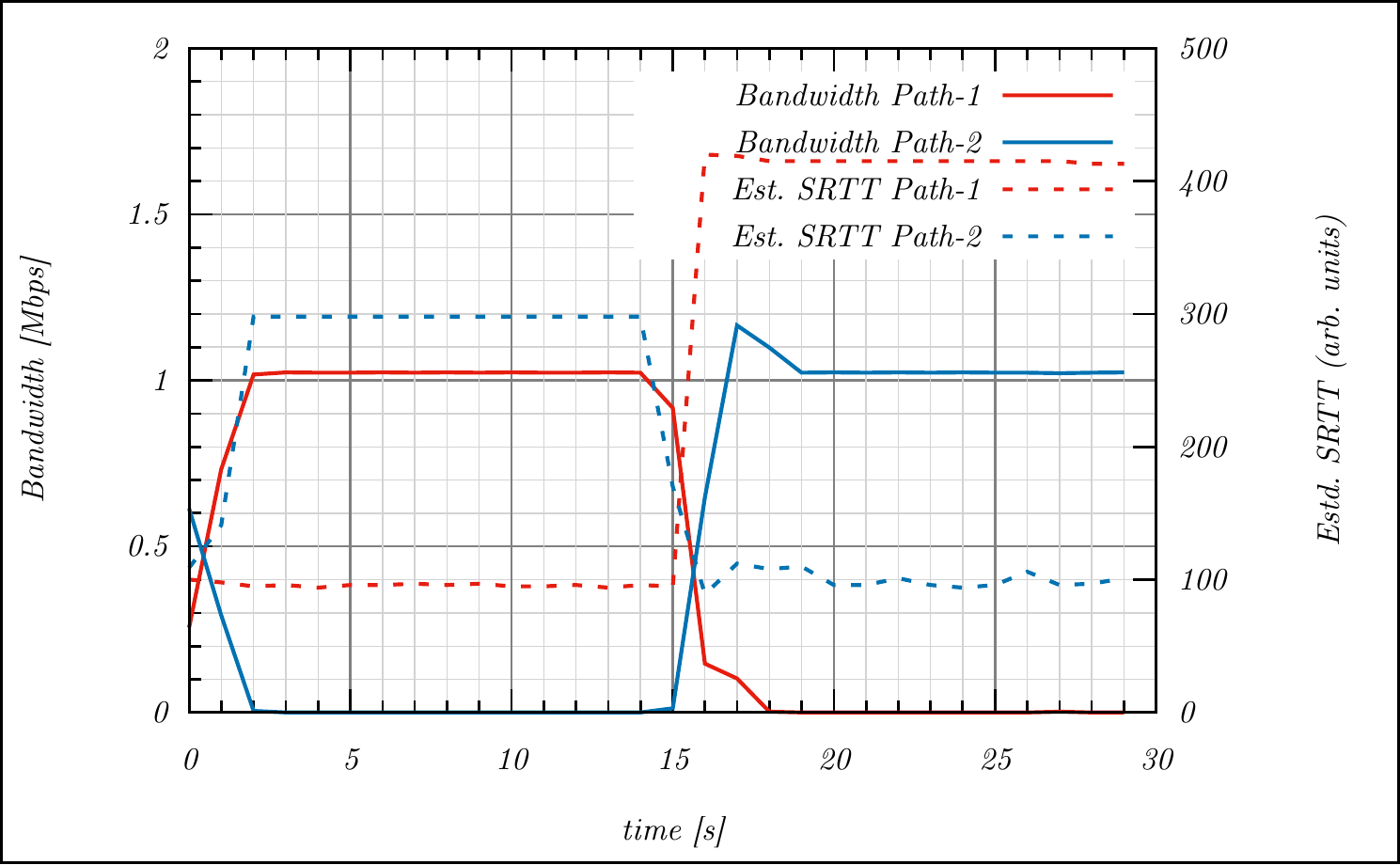}
\caption{Handover between network paths using the SRTT scheduler}
\label{fig:SRTT}
\end{figure}

Fig. \ref{fig:SRTT} shows the bandwidths measured on both network paths, as well as the estimated SRTT of each path as reported by the congestion control algorithm. During the first two seconds of the measurement, when the congestion windows limit the capacity of the flows, the scheduler uses both paths to forward the incoming traffic. Once the congestion windows are established, the scheduler prefers the first path because of its lower estimated SRTT. At this point, the bandwidth demand can fully be satisfied by the first path, and no traffic is sent over the second path anymore. Once the SRTT has been increased from 10 ms to 100 ms on the first path, the scheduler reacts and switches all traffic to path 2. This behavior matches the theory of operation; the SRTT scheduler seems to be able to correctly identify and select the path with the lowest SRTT.

The major advantage of this technique is that it does not introduce additional overhead; however, it is only effective as long as there is traffic on the network, whereas latency changes on unused paths remain undetected. This is easily noticeable in Fig. \ref{fig:SRTT} where the SRTT over path 2 (blue dotted line) remains at a high value up until around 15 seconds when new traffic is again transmitted over the path. 


The OTIAS algorithm has the largest effect in scenarios where there is a significant asymmetry between flows in terms of latency. OTIAS is able to build up large queues; the resulting latency increase is used in an effort to converge the path latencies. The scheduler has been examined in terms of buffer usage and SRTT estimation at the sender, which allows to verify the correct scheduler operation. Furthermore, the resulting packet scrambling is assessed at the receiver. This parameter is of special interest, since it is one major indicator of the effectiveness of the OTIAS algorithm.

In this experiment, the two network paths each provide 1 Mbps of bandwidth; their latency was adjusted to a 1:5 ratio (10ms on path 1 / 50 ms on path 2). In the first experiment, the OTIAS scheduler has been evaluated in a scenario where the test traffic bandwidth stays within the aggregated capacity of the two network links. A test bandwidth of 1.5 Mbps has been generated using the iperf network test tool. According to the theoretical predictions, OTIAS should be able to significantly reduce packet scrambling in this scenario. The measurement is then repeated using the SRTT scheduler for comparison.

\begin{figure}[ht!] 
\centering
\includegraphics[width=3.3in]{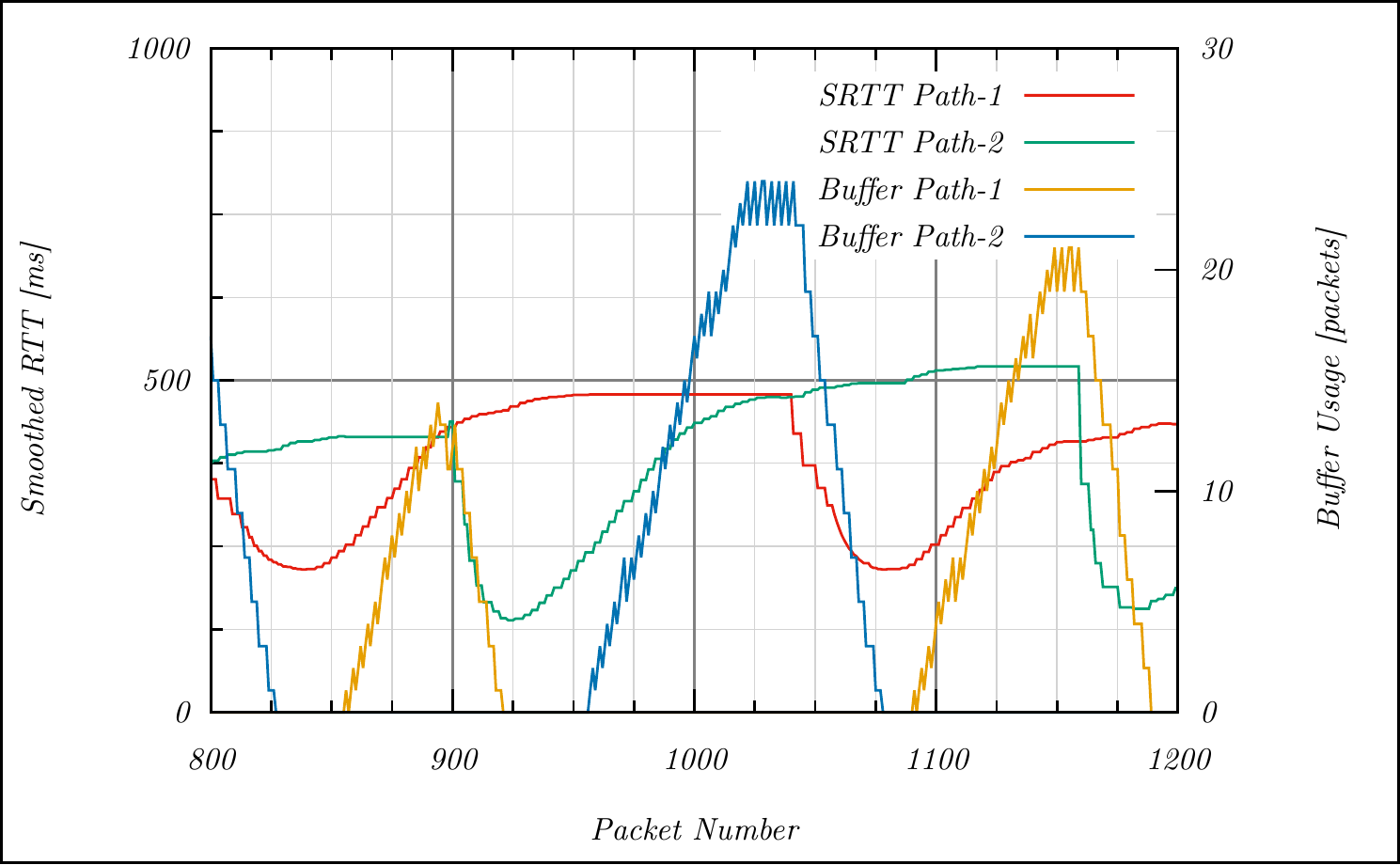}
\caption{Internal parameters of OTIAS in a moderately loaded environment}
\label{fig:otias1}
\end{figure}

Fig. \ref{fig:otias1} illustrates the behavior of the OTIAS algorithm, once the initial slow start phase is completed. There is a constant pattern of alternating buffer bloat and SRTT oscillation, which repeats throughout the entire measurement; the selected region of packets reflects this pattern. The scheduler chooses the path that provides the shortest overall latency (which is Path-1 for the packet range [800-900]); as a single path cannot satisfy the entire bandwidth demand, a queue builds up, which gradually increases the estimated SRTT (see Buffer Path-1 and SRTT Path-1 [850-890]). Once the estimated SRTT of Path-1 exceeds that of Path-2, the scheduler switches to the second path, which allows the buffer on Path-1 to drain (see Buffer Path-1 [900-920]). Since Path-2 cannot satisfy the bandwidth demand either, a buffer builds up (see Buffer Path-2 and SRTT Path-2 [960-1040]), and the cycle repeats. As new path measurements on the 'slower' path are missing, (as this path has not been used for some time), this contributes to buffer bloat on the 'faster' path. It is interesting to note here that the performance of an algorithm like OTIAS is dependent on the freshness of the path measurement information. The DCCP congestion control does not provide immediate path measurement information, but relies on application data and requires adaptation time. Modifying the scheduling module to deal with such problems is an important challenge for future work. 

\begin{figure}[ht!] 
\centering
\includegraphics[width=3.3in]{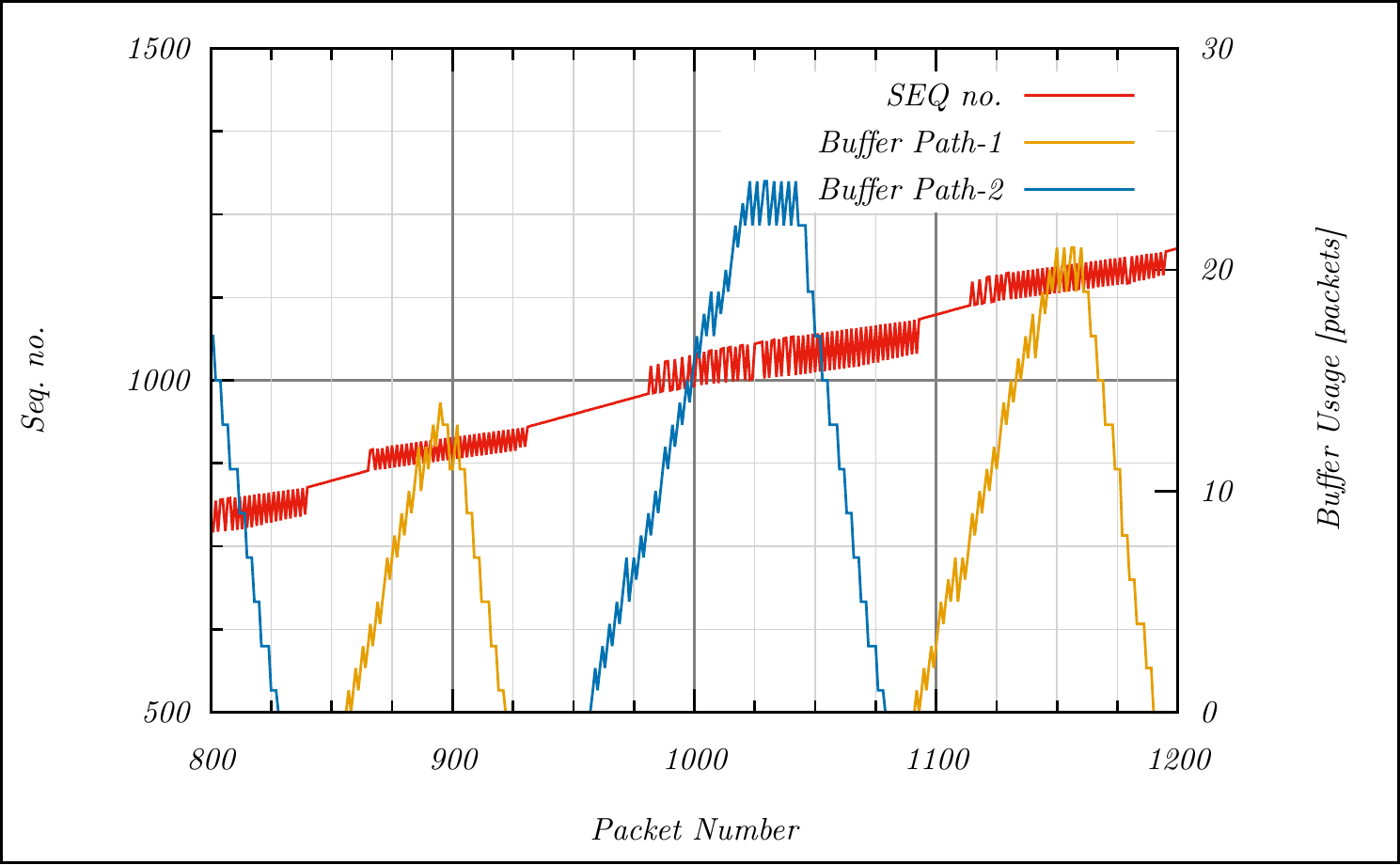}
\caption{Packet scrambling at the receiver when using OTIAS in a moderately
loaded environment}
\label{fig:otias2}
\end{figure}

During the first experiment, the packet scrambling has been recorded at the receiver. Fig. \ref{fig:otias2} presents the order of arrival of the sequence numbers assigned by the iperf tool. When correlated with the buffer bloat at the sender, a relation between the two properties becomes apparent; overloading a path seems to be a direct cause of packet scrambling at the receiver.

Overall, the OTIAS scheduler seems to be more well-suited for the heavily loaded network environments. In a low-load situation, the scheduler experienced issues such as periodic scrambling at the receiver, which seemed to correlate to the buffer bloat at the sender. These issues can be partially attributed to the imperfect channel estimation performed by the DCCP congestion control mechanism, as well as irregular buffer bloat in the network simulation hosts of the experimental testbed.  In a saturated network however, the scheduler was able to reduce the packet scrambling significantly and operated as designed.


\subsubsection{Performance Evaluation - Packet Reordering}

Operation of packet reordering is validated and analysed using the following scenario (using the architecture from Fig. \ref{fig:testbed-architecture}): path 1 has latency of 10ms, and on path 2, the latency changes during the experiment from the initial latency of 0ms to a latency of 40ms. 1Mbps UDP traffic is used for the experiments, with no additional packet loss. Round robin scheduling is used, to ensure the existence of out-of-order packets. At the reordering side,  adaptive timing threshold is used. 

\begin{figure}[ht!] 
\centering
\includegraphics[width=3.3in]{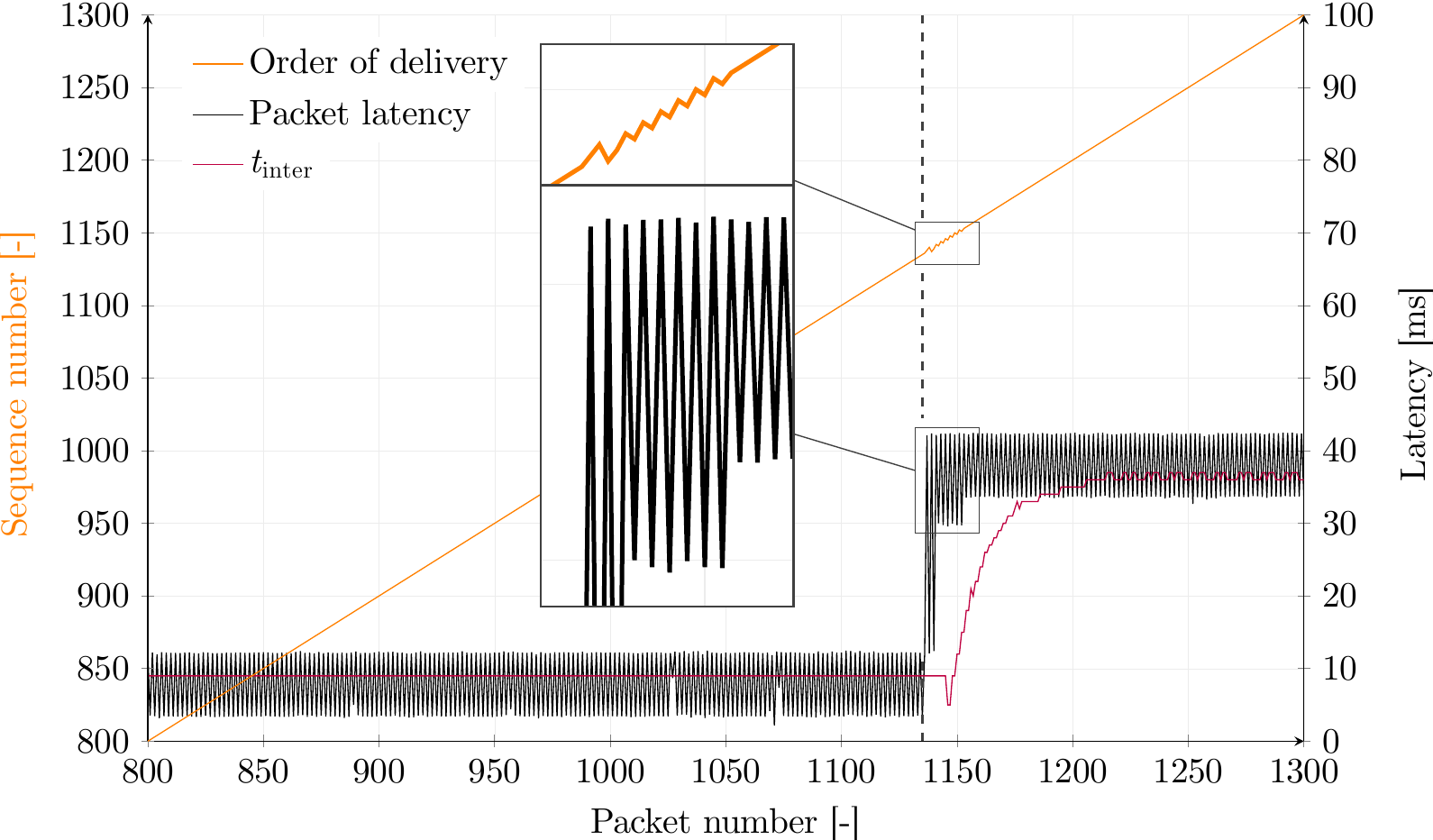}
\caption{Order of delivery for adaptive active reordering, heterogeneous scenario with a delay jump from \(\Delta t = 10ms\) to \(\Delta t = 30 ms\)}
\label{fig:reordering}
\end{figure}

The results of the experiment can be found in Fig. \ref{fig:reordering}. We can see from these results that adaptive solution is capable of handling such a change in latency. Before and after the latency change the solution forwards all packets in-order as can be seen from the perfect straight line for the order of delivery. The adaptation takes some time, because it is based on the measured delay values which rely on the DCCP congestion control operation.

\subsubsection{Managing Delay Variation}

The previous section demonstrated the use of packet scheduling and packet reordering algorithms within the multipath framework. The in-order packet delivery provided by these algorithms is particularly important for live and streaming multimedia applications. Currently, RTP or application-specific packet reordering mechanisms are deployed to ensure packet delay loss is minimized. However, we know that in multipath scenarios solutions based on RTP cannot efficiently operate over multiple paths because RTP operates at the media session level and not at the transport level, i.e., the receiver reports per media (audio or video)\cite{c2}. In RTP the receiver calculates the optimum playout using a low-point or mid-point averaging algorithm \cite{c29}. The algorithm averages over a large window size to reduce the impact of the few packets that get delayed. In a multipath scenario, the packets may be scheduled on paths with widely varying latencies and the above method may prove to be inadequate and insufficient.

As lot of multimedia traffic is carried by TCP today, MP-TCP could be applied to media streaming but it is worth noting that MP-TCP does not consider real-time data and diverse paths may lead to very long loss recovery delays.


We argue that a multipath solution based on MP-DCCP is able to handle packet latency variation quicker, delivering multimedia packet stream in-order without packet retransmissions, while preserving the increased throughput and communication reliability of the multipath delivery. To demonstrate this, a number of experiments have been performed on the network testbed described earlier in the paper. 

\begin{figure}[ht!] 
\centering
\includegraphics[height=2.3in]{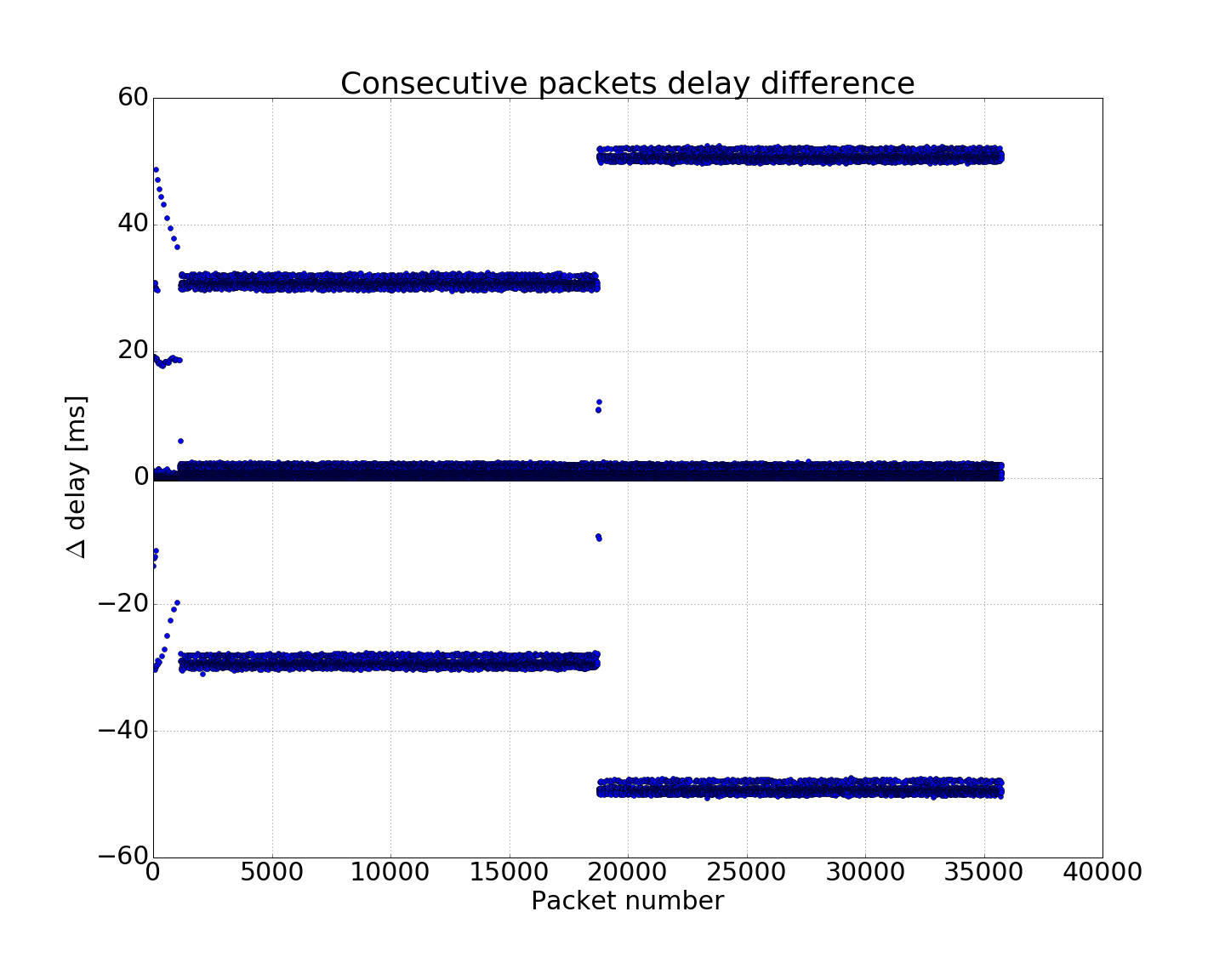}
\caption{Packet latency variation for the default 80/20 fixed ratio scheduling}
\label{fig:jitter-default}
\end{figure}

To evaluate the worst-case packet latency variation that will be experienced in a multipath scenario using the presented framework, in the first experiment it was assumed that 80 percent of the traffic will pass through the 'faster' path and 20 percent will pass through the 'slower' path. At some point during the experiment, the latency of the 'slower' path was increased by 30ms. The delay variation results are presented in Fig. \ref{fig:jitter-default}. The results show that - while the majority of the packets will arrive with minimal delay variation (packets grouped around 0 on y-axis in Fig. \ref{fig:jitter-default}) - a significant number of packets have been received out-of-order, with delay variation taking both positive and negative values, depending on which path the packets have taken. We measure the packet delay variation as the difference in the arrival time of two consecutive packets. This value is negative when a packet is required to wait for the packet preceding it in the sequence. If no packet reordering is applied, the network applications would have to deal with this variation. 

\begin{figure}[ht!] 
\centering
\includegraphics[height=2.3in]{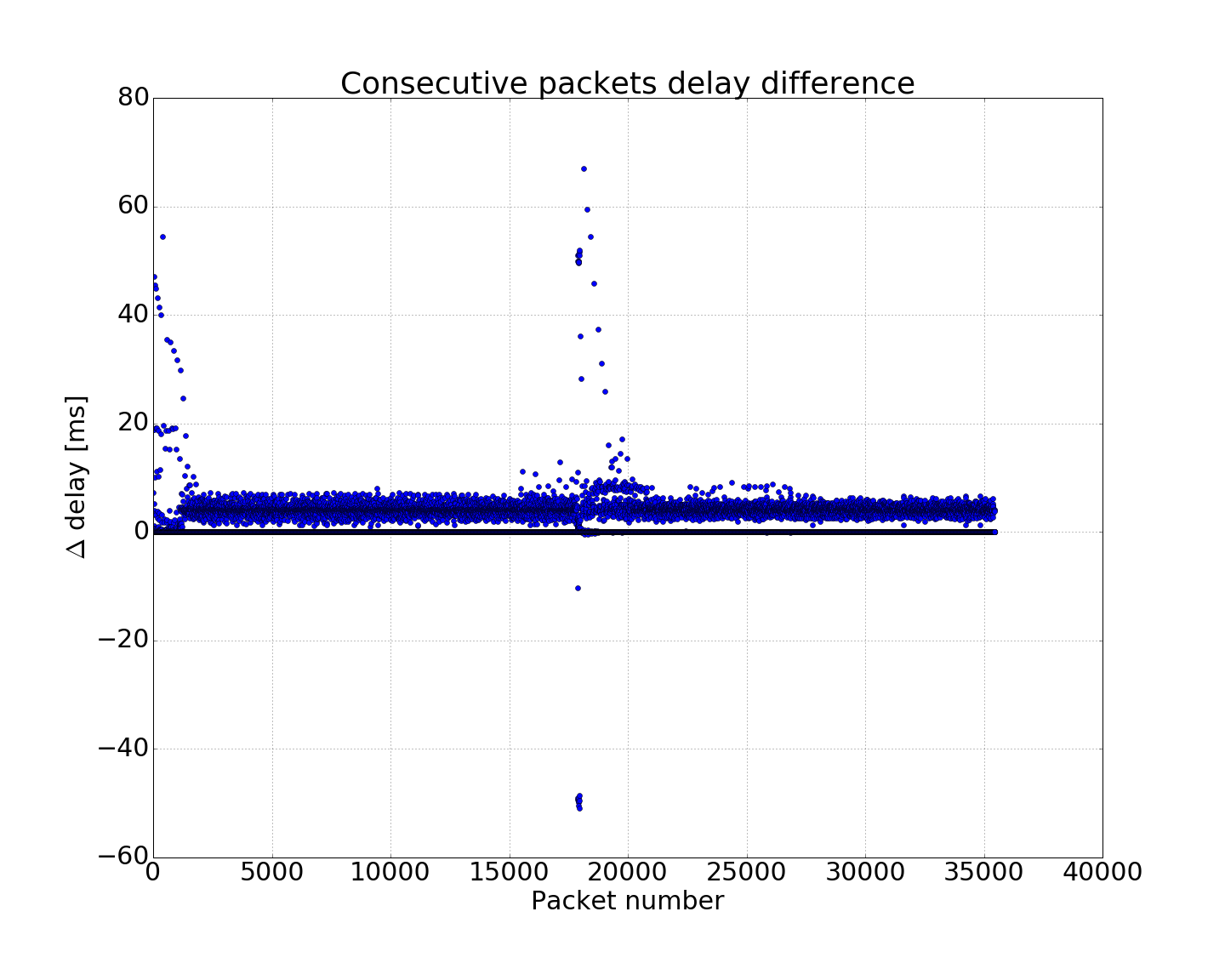}
\caption{Packet latency variation when adaptive reordering algorithm is applied}
\label{fig:jitter-reordering}
\end{figure}

\begin{figure}[ht!] 
\centering
\includegraphics[height=2.3in]{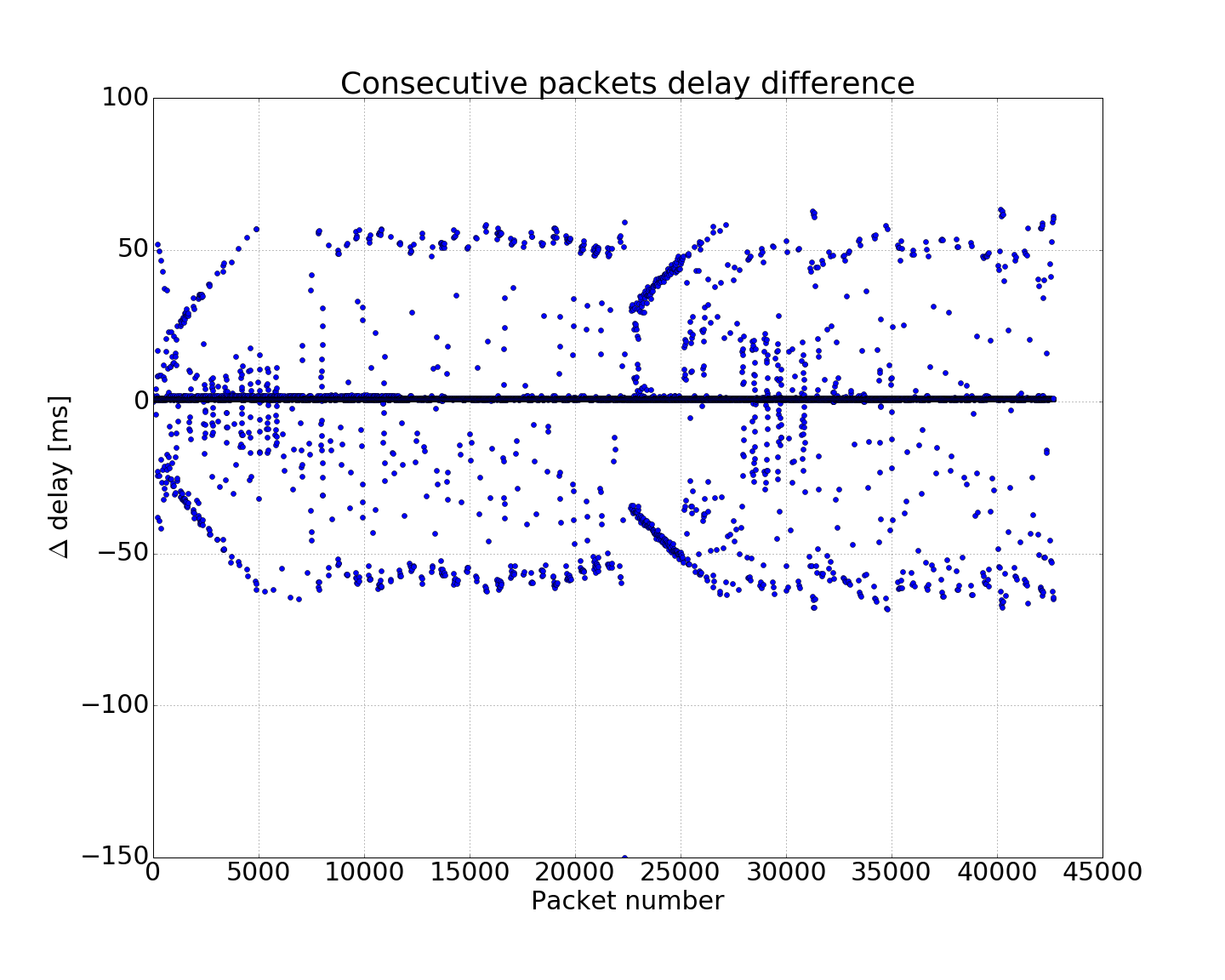}
\caption{Packet latency variation when OTIAS scheduling algorithm is applied}
\label{fig:jitter-otias}
\end{figure}

We then conducted experiments with scheduling and reordering algorithms deployed in the same scenario. Fig. \ref{fig:jitter-reordering} shows the packet delay variation for the same scenario when reordering with adaptive timing threshold is used. It can be seen from the scattering of the delay variation values that reordering significantly reduces the number of packets with a large delay variation. In addition to this, we can see that adaptive reordering handles the sudden change in path latency very well, by adapting to the new path latency quickly. 

\begin{figure}[ht!] 
\centering
\includegraphics[height=2.3in]{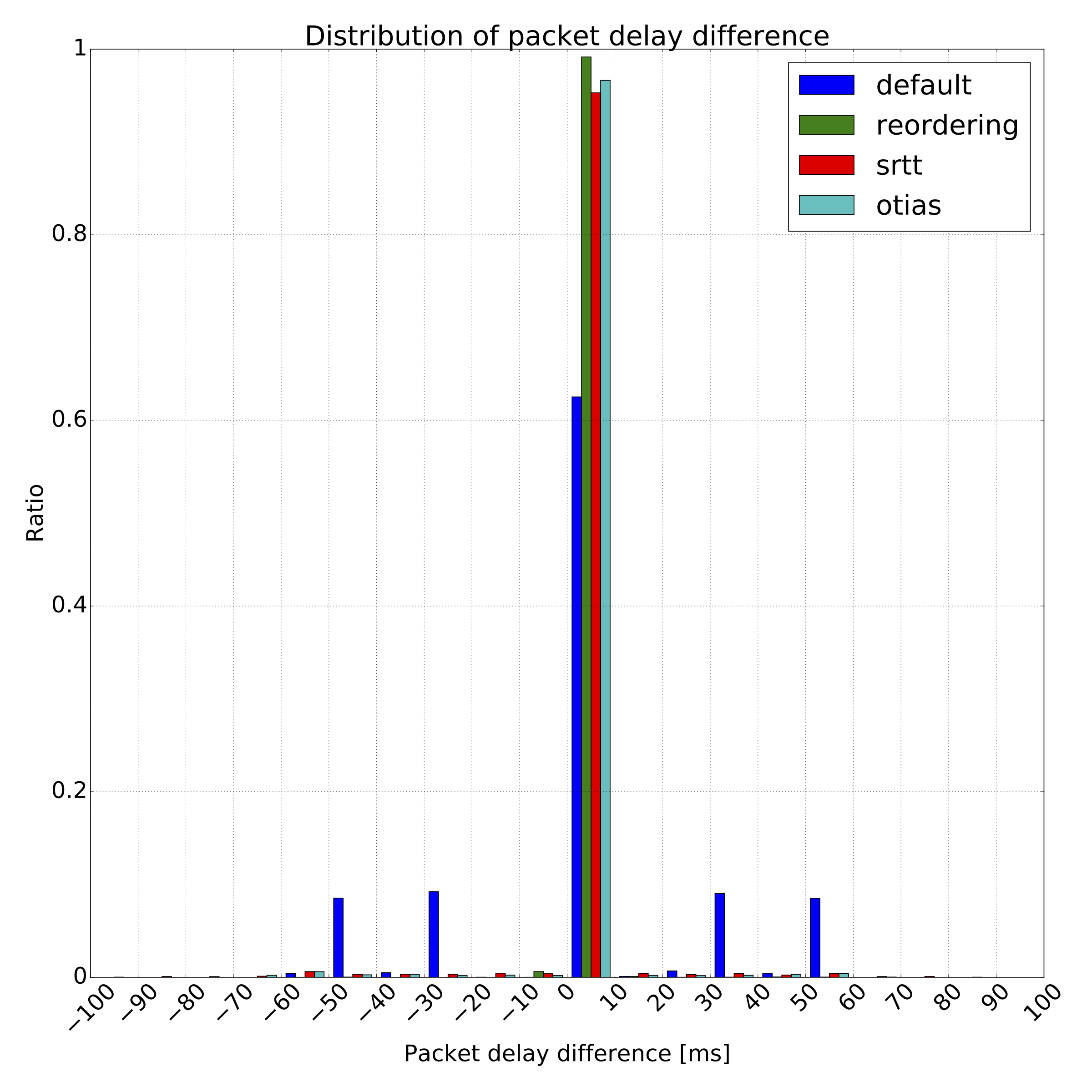}
\caption{Comparison of the distribution of packet latency variations}
\label{fig:jitter-all}
\end{figure}

Fig. \ref{fig:jitter-otias} shows the results of the same experiment if OTIAS scheduling was deployed. Comparing with the results in Fig. \ref{fig:jitter-default}, it is possible to see that packet scheduling has greatly reduced the number of packets that have been received out-of-order. There is a scattered quantity of packets with large delay variation, but the great majority will have minimal variation. The results when SRTT scheduler is used are very similar.

Aggregated results from the experiments are given in Fig. \ref{fig:jitter-all}, where the benefit of packet scheduling/reordering within the multipath framework can be seen clearly. The values of the packet delay variation if OTIAS packet scheduling, or SRTT packet scheduling or adaptive packet reordering are deployed are all concentrated around 0, whereas the values for the case without any packet processing (the 'default' case) vary significantly. This shows that the multipath framework is able to remove the packet latency variation and provide in-order delivery of the packets to the applications.

\section{CONCLUSIONS}

In this paper, we presented a new IP compatible multipath framework for heterogeneous network access. The framework uses Multipath DCCP protocol to provide support for multipath traffic delivery. MP-DCCP obtains path characteristics information using internal congestion control mechanisms and may use this information to distribute the traffic  between different paths. In the paper, we describe the protocol design of MP-DCCP and use simulation and testbed experimental results to motivate and justify the multipath framework design and the use of MP-DCCP. Simulation results show the increase in the throughput of UDP traffic in the framework, and the advantage of using DCCP tunnels, and experimental results show the benefits of packet scheduling and reordering algorithms, particularly in terms of the improved packet latency variation for UDP traffic.
This work continues. In the future, we will run a large scale experimental evaluation with more realistic path latency variances, inclusion of packet loss in the experiments, and testing the performance of real services and their interaction with MP-DCCP. 


\addtolength{\textheight}{-12cm}   
\end{document}